\documentclass[aps,amssymb,amsmath,nofootinbib,notitlepage,superscriptaddress,twocolumn]{revtex4-1}

\usepackage{amsfonts}
\usepackage{graphicx,graphics,epsfig,times,bm,bbm,mathrsfs}
\usepackage{subcaption}
\usepackage{comment}
\usepackage{caption}
\usepackage{tikz}
\usepackage{lipsum} 
\usepackage{bm}
\usepackage{setspace}
\usepackage{amsmath}
\usepackage{graphicx}
\usepackage[nearskip,margin = 0pt]{subfig}
\usepackage{braket}

\begin{document}

\title{Building a global quantum internet using a satellite constellation with inter-satellite links}

\author{Alireza Shabani}

\affiliation{Center for Quantum Networks, University of Arizona, Tucson, Arizona 85721, USA}

\date{\today}

\begin{abstract}

The quantum internet is a global network to distribute entanglement and communicate quantum information with applications in cybersecurity, quantum computing, and quantum sensing. Here, we propose building a quantum internet using a constellation of low-Earth-orbit satellites equipped with inter-satellite laser links. Our proposal is based on the downlink model of photon transmission, where satellites carry entangled photon sources and/or have a mirror relay system to redirect the photon path in space. We show that few MHz entanglement distribution rates are possible between the US, Europe and Asia, with a multiplexed entangled-photon source generating pairs at 10 GHz. Our proposal demonstrates the importance of passive optics in realizing a satellite-based quantum internet, which reduces dependency on quantum memory and repeater technologies.

\end{abstract}

\maketitle

\section{Introduction}

Quantum networking is one of the main pillars of quantum technology and an enabler of quantum cryptography, quantum computing, and quantum sensing \cite{wei2022towards,simon2017towards}. The pinnacle of quantum networking is the creation of the quantum internet, a global-scale network to distribute quantum resources and exchange quantum information \cite{kimble2008quantum,cacciapuoti2019quantum,van2022quantum}. Quantum internet will work in synergy with the classical internet we have today, enabling users to securely generate, access, send, and receive quantum information for tasks such as key encryption for quantum-secure communications \cite{gisin2007quantum,yuan2010entangled,xu2020secure,steiner2023continuous}, blind quantum computing \cite{fitzsimons2017private,li2021quantum}, or distributed quantum sensing \cite{gottesman2012,padilla2025}. Quantum networks rely on optical communication where information is transmitted as optical qubits \cite{moody20222022,wang2020integrated,boileau2004robust,chapman2022hyperentangled}. Quantum networks can leverage the existing optical classical communication technologies. However, because of the nature of quantum signals, and no-cloning property \cite{Wootters}, they cannot employ any signal amplification as used for transmitting classical communications over long distances. For example, the main obstacle for fiber-based networks is overcoming the $\sim$0.2 dB/km loss, which limits the network channel range to at most $\sim100$ km. This range could be extended to regional- or continental-scale using quantum repeaters \cite{azuma2023quantum,li2019experimental}, or ground-based vacuum beam guides \cite{huang2023vacuum}; however, these solutions rely on new technologies that require overcoming several challenging hurdles for scaling, also difficult to deploy under the oceans. Quantum repeaters, for example, perform quantum error correction and therefore have the same challenges of developing an quantum computer \cite{li2022photonic,niu2023all,yang2023asynchronous}. It is thus critical to develop a novel approach that is capable of generating and distributing quantum signals over global-scale distances whilst utilizing existing technologies and techniques to the greatest extent possible.

\begin{figure}[b]
    \vspace{-10pt}
  \includegraphics[width=0.9\linewidth]{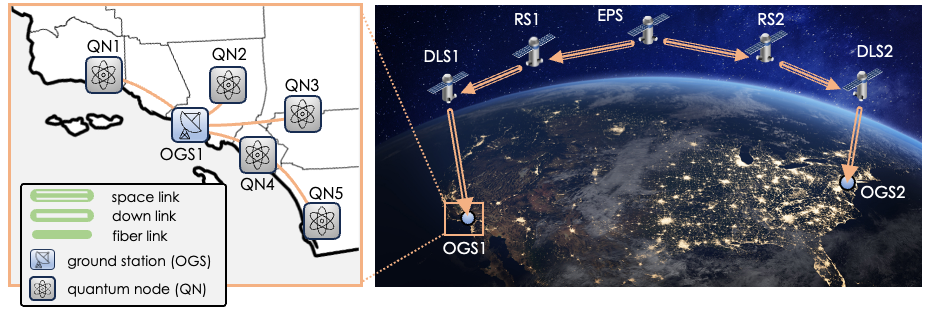}
  \caption{A conceptual overview of a space-based quantum network comprising a satellite constellation. An entangled-photon pair satellite (EPS) produces entangled-pairs. The EPS distributes quantum information with low loss across 1000's of kilometers using relay satellites (RS1 and RS2) and down-link satellites (DLS1 and DLS2) to transfer the information to optical ground stations (OGS1 and OGS2) with high capacity (above MQbits/sec) from Los Angeles to New York. At the OGS, local regional networks comprise, fiber or free-space quantum channels for 10 km distances to distribute quantum information to quantum nodes.}
  \label{fig:intro}
\end{figure}

Quantum communication via satellites is a promising path to realize a global quantum internet.  Satellite-based entanglement distribution over $\sim1200$ km has already been demonstrated \cite{yin2017satellite} and more experimental projects are under development \cite{rivera2024,hutterer2022, JENNEWEIN2025, lim2020}. Here, we propose a repeater-less, space-based quantum network that can serve as the backbone for a global quantum internet.  Illustrated in Fig. \ref{fig:intro}, the approach relies on a constellation of low-Earth-orbit (LEO) satellites equipped with inter-satellite laser links (ISLL), which only experiences diffraction losses through space transmission and $0.5\%-5\%$ loss at each satellite link.

\onecolumngrid

\newpage

\twocolumngrid

  Three different classes of satellites provide the necessary functionality for global-scale networking: an entangled-photon pair satellite (EPS) for the generation of quantum signals, relay satellites (RS) with an optical sub-system of a few ultra-low-loss mirrors (up to $99.9\%$ reflection) to redirect and route signals, and down-link satellites (DLS) to distribute signals to optical ground stations (OGSs). The satellite constellation relies on reflective optics, as opposed to refractive optics \cite{goswami2023satellite}, to avoid chromatic aberrations and enable broadband operation for, e.g. spectral multiplexing with a variety of wavelengths spanning visible to mid-infrared. Here we consider polarization-entangled qubits, which are generally more immune to atmospheric turbulence and diffraction for high-fidelity entanglement distribution compared to other encodings, although some developments in time-bin encoding with have been recently reported \cite{jin2019genuine,vallone2016interference,steiner2023continuous}. Polarization encoding is also motivated in part by recent advances in spontaneous parametric down-conversion (SPDC) with both integrated photonic entanglement sources \cite{baboux2023nonlinear,wang2016chip,schuhmann2023hybrid} and bulk nonlinear optics, the latter of which can produce polarization-entangled pairs at $>1$ GHz rates \cite{cao2018bell} with compact bulk optical setups. We show that with a 10 GHz source, entanglement distribution rates of few MHz are possible between international OGSs, which can further distribute signals regionally ($<10$ km) through fiber-based communications links to local quantum nodes (QNs).

\section{Modeling Space-Based Quantum Communication Links}

\begin{figure}[h]
  {\includegraphics[width=0.4\textwidth]{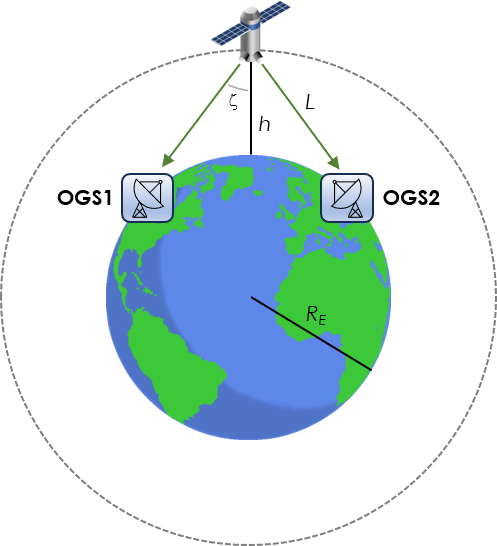}}
  {\caption{\label{fig:single_sat} A single satellite at altitude $h$ distributing entangled photon pairs between two OGSs with a channel length \textit{L}. The earth radius is denoted by $R_E$, and the angle between the zenith and the OGS is denoted by $\zeta$.}}
\end{figure}

There is a large body of work on modeling satellite quantum communication for a single satellite \cite{bourgoin2013,abasifard2023ideal, meister2024}, or a constellation of independently operating satellites with or without quantum repeaters \cite{boone2015, khatri2021spooky, gundougan2021proposal, liorni2021}. Here, we consider a constellation with ISLLs for quantum networking. The satellite constellation can orbit at different altitudes depending on the application: Low Earth Orbits (LEOs) range from 160 to 2,000 km in altitude, Medium Earth Orbits (MEOs) range from 2,000 to 35,786 km, and the Geostationary Equatorial Orbit (GEO) is at 35,786 km.  GEO orbits offer the advantage of wider coverage from a single satellite but suffer from high photon loss, whereas LEO orbits experience lower loss but have narrower coverage. Our proposal emphasizes a reduction in total channel loss and thus focuses on LEO at a 500 km altitude, where satellites, traveling at 7.62 km/s, orbit Earth 15.25 times in 24 hours. Several important design considerations impact quantum communication rates, including diffraction, atmospheric absorption and scattering, relay satellite transmission and throughput, the operation wavelength, and the quantum light source design. Here, we consider each of these below for a single satellite in an LEO orbit at an altitude $h$, as shown in Fig. (\ref{fig:single_sat}).

\vspace{5pt}

\noindent {\it{Diffraction Loss---}} We assume the satellite resides at or above the Karman line of 100 km above the Earth for which the atmosphere is considered to be absent. We first consider diffraction due to transmission in an almost vacuum space. A photon with wavelength $\lambda$ leaves a telescope of a satellite with beam waist $w_0$ and is received by a satellite telescope with aperture $r_a$ after traveling a distance $L$ in a vacuum. With negligible atmospheric loss, the transmittance is given by 

\vspace{-10pt}
\begin{equation}
\eta_{\rm{fs}}(L)=1-\exp(-\frac{2r_a^2}{w_0^2[1+(L/d_R)^2]}),
\label{freespace_trans}
\end{equation}

\noindent where $d_R=\pi w_0^2/\lambda$ is the Rayleigh range. The receiving telescope can be an optical ground station (OGS) or a relay satellite. Here, we consider a relay satellite with negligible propagation loss other than diffraction. In the regime of $L>>d_R$ and $r_a<<w_0L/d_R$, the transmittance depends quadratically on the distance

\vspace{-10pt}
\begin{equation}
\eta_{\rm{fs}}(L)\approx \frac{2r_a^2}{w_0^2(L/d_R)^2}=\left(\frac{\sqrt{2}\pi r_a w_0}{L \lambda}\right)^2.
\label{freespace_trans2}
\end{equation}

\noindent For satellite-to-ground, upon entering the atmosphere, photons experience additional loss due to absorption, scattering, air turbulence, and potentially complete opacity due to extreme weather and clouds. We discuss technical solutions for the effect of turbulence and clouds in the discussion section. 

\vspace{5pt}

\noindent{\it{Atmospheric Loss---}} Here we use the Beer-Lambert law for a homogeneous absorptive layer \cite{khatri2021spooky}. For small zenith angles, the atmosphere transmittance $\eta_{\rm{atm}}$ is given by

\begin{equation}
\eta_{\rm{atm}}(L,h)= \begin{cases}
                    \eta_0^{\sec\zeta},    & \text{if $-\frac{\pi}{2}<\zeta<\frac{\pi}{2}$}\\
                    0 & \text{if $|\zeta|\geq\frac{\pi}{2}$}
                    \end{cases} \\
\end{equation}
The constant $\eta_0$ represents the atmospheric transmittance at zenith. We use $\eta_0=0.47$ at wavelength of 810 nm \cite{bourgoin2013}. The zenith angle $\zeta$ is given by

\begin{equation}
\cos \zeta =\frac{h}{L}-\frac{1}{2}\frac{L^2-h^2}{L R_E}
\end{equation}

\begin{figure}[t]
  \includegraphics[width=1\linewidth]{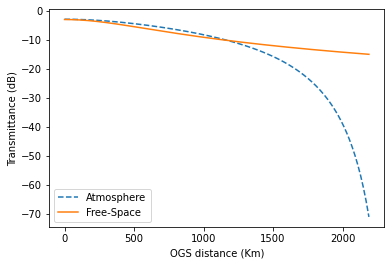}
  \caption{Transmittance as a function of distance between OGS and satellite nadir for diffraction losses in vacuum (orange solid curve) and absorptive losses in the atmosphere (blue dashed curve), using a satellite altitude $h=500$ km.}
  \label{fig:loss_functions}
\end{figure}

Fig. \ref{fig:loss_functions} shows the transmittance as a function of the distance between the OGS and the satellite nadir for photons traveling through a vacuum (diffraction only) and through the atmosphere (absorption only). We consider a Gaussian mode with a beam waist of 10 cm, a wavelength of 810 nm (see the discussion), and a receiver telescope aperture of 75 cm. Up to 1000 km, both mechanisms result in approximately 10 dB loss. Above approximately 1500 km, the atmospheric transmittance drops precipitously. Thus, we use 1000 km as a critical design point—satellite-to-OGS communication would only occur when a satellite is within a 1000 km vicinity of the OGS to avoid significant atmospheric absorption losses.

\vspace{5pt}

\begin{figure}[b]
  \includegraphics[width=1\linewidth]{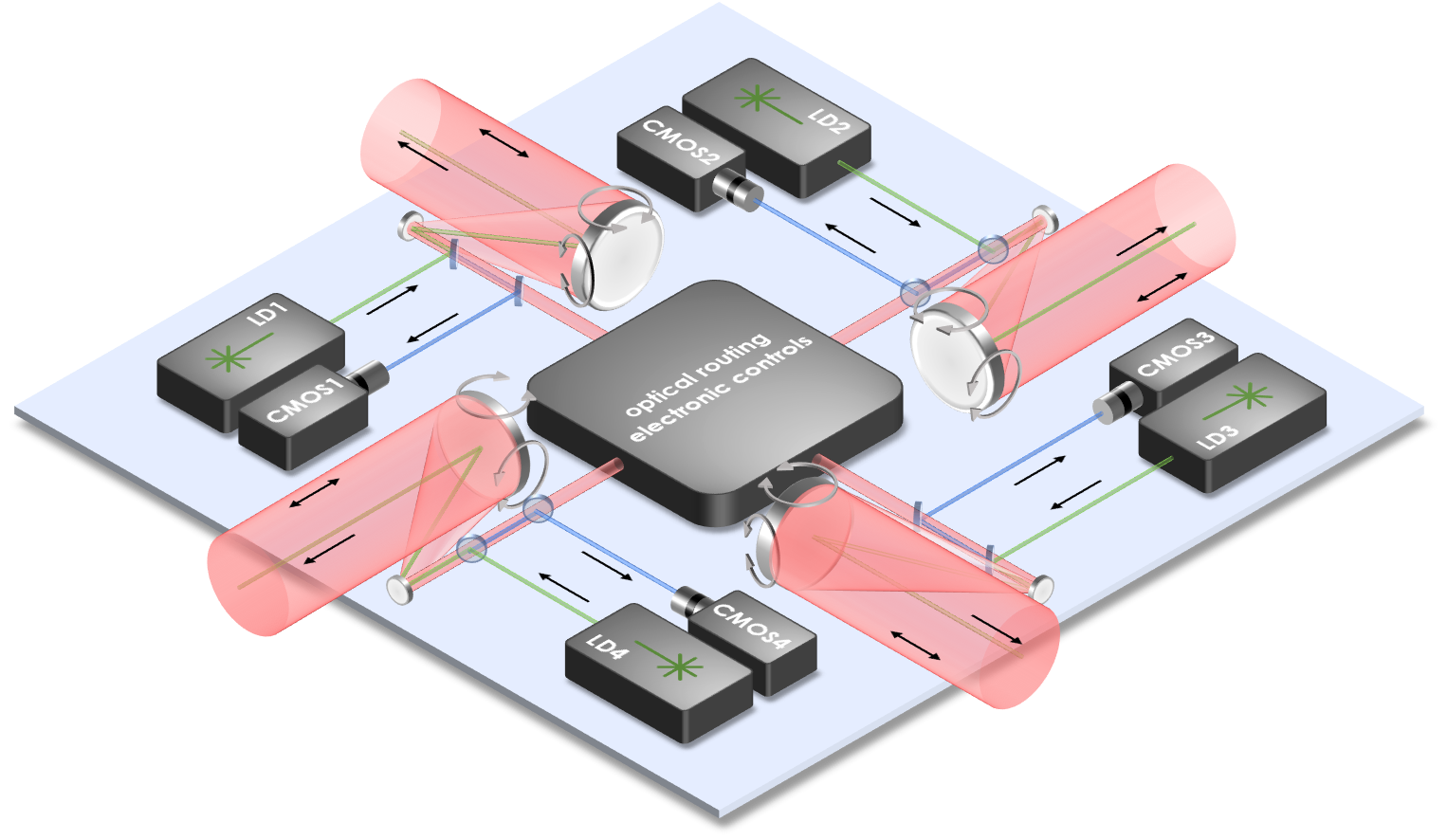}
  \caption{A schematic illustration of a relay satellite comprising a bifocal mirror system with two-axis gimbal mirrors for alignment and tracking. Each relay satellite is equipped with a single laser diode (LD) (or series of LDs depending on wavelength and power requirements) and CMOS cameras to enable PAT protocols for ISLLs. An optical switching system comprising flip-mount mirrors and electronic controls enables routing between channels. Dichroic mirrors are used to separate the LD output and CMOS detection input laser signals. With currently available mirror materials and dielectric coatings, $>99\%$ throughput from port-to-port is possible.}
  \label{fig:relay}
\end{figure}

\noindent{\it{Relay Satellite---}}The concept of reflecting light by a mirror in orbit was successfully demonstrated in 1990, known as the relay mirror experiment \cite{begley1996proposed}. In this experiment, a 61-cm-diameter mirror was launched into a LEO at 450 km altitude, and the mirror was controlled by fine tracking of two cooperative beacons. In our model for a relay satellite (RS), we envision a system of two or four independently controllable, but optically coupled, telescopes in a single RS (see Fig. \ref{fig:relay}). A two-port relay satellite has been previously built and tested in the lab, motivated in part by the 1990 experiment \cite{spencer1996}. For each port of the relay, a six-mirror optical system is needed, including two primary and two secondary telescope mirrors and two routing mirrors between the telescopes. As shown in Fig. \ref{fig:relay}, a four-port RS could be controlled through an internal switching system to route photons along the ideal constellation path to minimize loss and maximize the transmission rate between any two optical ground stations depending on location and atmospheric conditions. An integrated pointing, acquisition, and tracking (PAT) system comprising laser diodes and CMOS imaging with on-board controls and processing would allow for fine tracking and alignment between neighboring RSs. Typical commercially telescope mirrors can be made from several different materials, including gold or silver ($>95\%$ reflection from optical wavelengths spanning 800 nm to 10 $\mu$m). Dielectric Bragg coatings can improve the reflection to $>99.9\%$ for a narrower wavelength range ($\sim800-1000$ nm). For a six-mirror, two-telescope RS (and similarly for EPS and DLS mirrors), we consider a total relay loss of $<1\%$ possible with dielectric mirrors. 

\vspace{5pt}

\noindent{\it{Operational Wavelength---}}When considering the operational wavelength for the satellite constellation, multiple factors must be considered. These include the entangled-pair source operational wavelength; transmissivity of the optical systems, sub-systems, and atmosphere; wavelength-dependent diffraction; adaptive-optical solutions for counteracting atmospheric turbulence; and day versus night operations. Considering these factors holistically, several recent studies suggest that $\sim800$ nm is an ideal operational wavelength that can yield the largest signal-to-noise ratio (SNR) that is equal to the channel performance $S_Q/S_B$, where $S_Q$ ($S_B$) is the probability to detect the desired signal photon (background event due to optical noise, solar irradiation, or detector dark counts) \cite{lanning2021quantum,abasifard2023ideal}. The satellite-to-OGS aperture-to-aperture coupling efficiency is $\sim4$ times larger for 775 nm than 1550 nm, whereas the atmospheric transmission is $\sim90\%$. Altogether, the SNR for 775-810 nm is $\sim2$ times larger and the total detected signal rate is nearly 4 times larger for the 800 nm range \cite{gruneisen2021adaptive}. Importantly, several high-rate and high-efficiency bulk nonlinear and integrated photonic entangled-photon pair sources exist for this wavelength range \cite{moody2020chip}, and optical systems and sub-systems (including lasers, optics, modulators, detectors, etc.) are commercially available. For our modeling discussed below, we consider $\lambda = 810$ nm.  

\vspace{5pt}

\noindent{\it{Entangled-Photon Pair Source---}}The majority of entangled-photon pair sources, including those commercially available, rely on spontaneous parametric down-conversion (SPDC) in nonlinear bulk crystals and compact, table-top optical setups \cite{steinlechner2012high,kwiat1995new}. Through SPDC in a nonlinear crystal or waveguide such as periodically poled KTP (PPKTP) \cite{cao2018bell} or lithium niobate (PPLN) \cite{sun2023degenerate,huang2022high}, 405 nm (775 nm) pump photons are converted through the $\chi^{\left(2\right)}$ nonlinear response to produce degenerate 810 nm (1550 nm) photon pairs, traditionally called the signal and idler. Using a continuous-wave laser with a center wavelength of 405 nm to pump a nonlinear crystal from both clockwise and counter-clockwise directions simultaneously in a Sagnac interferometer (see the inset to Fig. \ref{fig:intro}), down-converted photons at 810 nm can be generated with the biphoton wavefunction $\ket{\Psi}=\left(\ket{H}_S\ket{H}_I + \ket{V}_S\ket{V}_I\right)/\sqrt{2}$, where $H$ and $V$ denote horizontal and vertical polarization with respect to the interferometer reference frame. Such PPKTP and PPLN waveguide sources with type-0 phase matching can produce polarization-entangled photon-pairs with generation rates up to 1 GHz requiring 10's of milliwatts of pump power \cite{yin2020entanglement}. Bulk Sagnac interferometric sources can be designed as a compact, thermally and mechanically stable satellite sub-system with a single cubesat unit footprint. 

Recent experiments have also demonstrated both polarization and time-bin encoded qubits using integrated photonics with pair generation rates up to $\sim20$ GHz for milliwatt pump power for 800 nm and 1550 nm wavelength operation \cite{schuhmann2023hybrid,steiner2021ultrabright,baboux2023nonlinear,moody20222022}. Novel time-bin encoding protocols have been developed that can relax the hardware requirements and may simplify direct free-space channel communications, particularly in the presence of depolarizing channels such as metropolitan fiber or atmospheric turbulence \cite{jin2019genuine}. Integrated photonics for satellite-based quantum networking has not been explored to any significant extent, but the improved energy efficiency, compact footprint, and potential for massive multiplexing make this a promising direction for scalable high-rate satellite-based sources \cite{steiner2023continuous}. In this paper, we consider both 1 GHz and 10 GHz sources, where the latter can be achieved with high-power pumped bulk optics or via multiplexed photonic integrated circuit sources \cite{Moody2025}.

\section{A Single Satellite}

Before considering a full constellation, let us evaluate the performance of a single satellite for entanglement distribution. In our simulations, we consider the satellite has onboard a photon source generating entangled pairs at an $N_{EPS} = 1$ GHz rate. The rate of entangled pairs received at two OGSs are given by $\eta_{total} = N_{EPS}\left(\eta_{\rm{fs}}(L_1)\eta_{\rm{atm}}(L_1)\times
\eta_{\rm{fs}}(L_2)\eta_{\rm{atm}}(L_2)\right)$, where $L_{1,2}$ is the satellite distance from the OGS$_{1, 2}$. We consider the satellite emitting photons with a beam waist $w_0=10$ cm and OGSs have telescopes with an aperture $75$ cm.

Table I summaries the maximum and the average achievable rates between different cities in the United States. We have done a brute force optimization over the satellite inclination, RAAN, and altitude $h\in\{500,800,1000,2000\}$ km. Using these realistic parameters taken from previous or estimated satellite and OGS configurations, distributing entanglement at a few MHz rate is within reach for single satellite over a 1000 km distance. That confirms the possibly of building regional and state-wide quantum networks with a constellation of satellites without requiring ISLLs. 

\begin{table}[h!]
  \begin{center}
    \label{tab:table1}
    \footnotesize
    \begin{tabular}{l|l|l|l}
      \textbf{OGS1-OGS2 (kilometers)} & \textbf{max. rate} & \textbf{ave. rate} & \textbf{altitude}\\
      \hline
      Los Angeles-Santa Barbara (140)& 57.3 MHz & 5.3 MHz & 500 km\\
      Los Angeles-San Francisco (559)& 30.9 MHz & 4.5MHz & 500 km\\
      Chicago-New York (1149) & 6.9 MHz & 0.95 MHz & 500 km\\
      Los Angeles-New York (3936) & 25.5 kHz & 6.9 kHz & 2000 km\\
    \end{tabular}
  \end{center}
  \vspace{-10pt}
      \caption{Maximum and time-average rate of achievable entangled bits (ebits) distributed between U.S. cities with a single satellite. We assume a 1 GHz polarization-entangled pair source, a beam waist of 10 cm, and an OGS telescope aperture of 75 cm.}
\end{table}

\section{a satellite constellation with inter-satellite links}

Interest in developing satellite constellations to achieve global internet connectivity has surged over the past few years. Some advantages of satellite-based internet solutions over optical fiber include global coverage, rapid deployment, and reduced infrastructure costs. Although satellite constellations have been in operation for decades, only recently some constellations have been equipped with ISLLs. Starlink and Kuiper projects have reported 100 GHz ISLL rates over 1000 km, which highlights the maturity of the existing technology. The ISLLs for classical internet communications links, however, cannot be readily adapted for quantum networking, because the optical routers used on satellites are optoelectronic---similar to fiber optical communications. However, what is important is the achieved level of mechanical precision that is required to link two orbiting satellites. Our proposed scheme leverages the technological maturity and knowledge built over decades to build a quantum satellite constellation with ISLLs. 

\begin{figure}[t]
  \includegraphics[width=0.8\linewidth]{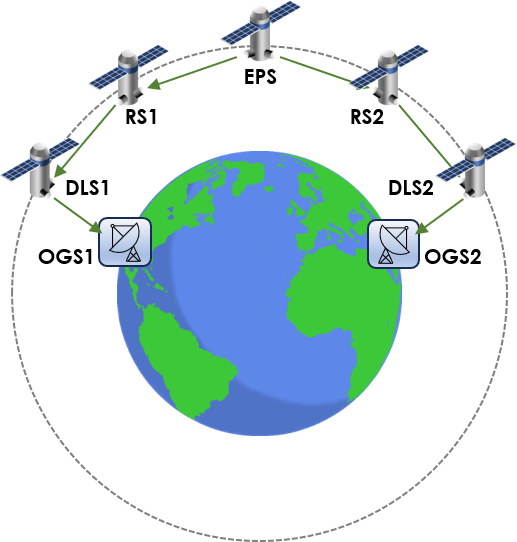}
  \caption{Distributing entanglement between two OGSs with a chain of relay satellites.}
  \label{fig:relay_chain}
\end{figure}

We first consider a single chain of satellites moving in the same orbit, as illustrated in Fig. \ref{fig:relay_chain}. To assess the limits of achievable entanglement distribution rates, we ask this question: given two OGSs (OGS1 and OGS2), what is the minimum number of relay satellites needed to create a path between the OGSs such that the last two relay satellites are directly on the OGS zenith? One constraint that must be satisfied is ensuring all ISLs stay above the atmosphere's edge at 100 km. We calculate the rates when the satellite emitting the entangled pairs is in the middle of the path, see Fig. \ref{fig:relay_chain}. 

\begin{figure}
\includegraphics[width=1\linewidth]{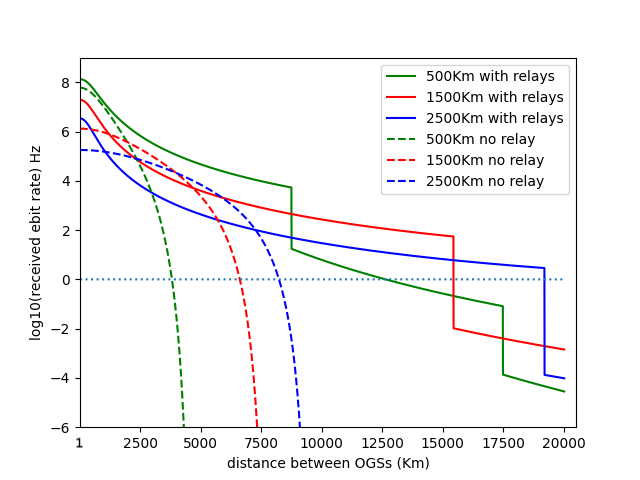}

\caption{Achievable rate of entanglement distribution for a chain of relay satellites as shown in Fig. \ref{fig:relay_chain}. Different colors correspond to different satellite altitudes. Dotted lines show the performance of a single satellite with no rely.}
\label{fig:Rates}
\end{figure}

Figure \ref{fig:Rates} summarizes the entanglement distribution rates for different satellite altitudes with and without relays (i.e., a single satellite and two OGSs) for comparison. The points of discontinuity refer to adding additional relay satellites when the optical link path drops below the 100 km Karman line threshold. The results are encouraging---with only six relay satellites, one can distribute entanglement at global scales at a few kHz ebit/sec with an entangled source emitting at 1 GHz (ebit = entangled-pair bit).

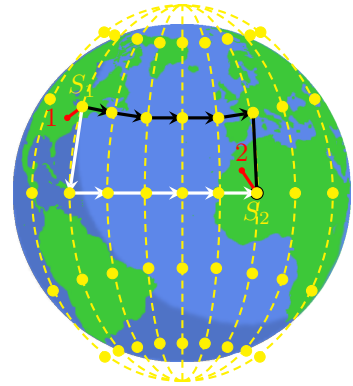
\begin{figure}
\centering
\label{fig:globe}
\begin{tikzpicture}
\node[inner sep=0pt] (russell) at (2,2)
    {\includegraphics[width=.25\textwidth]{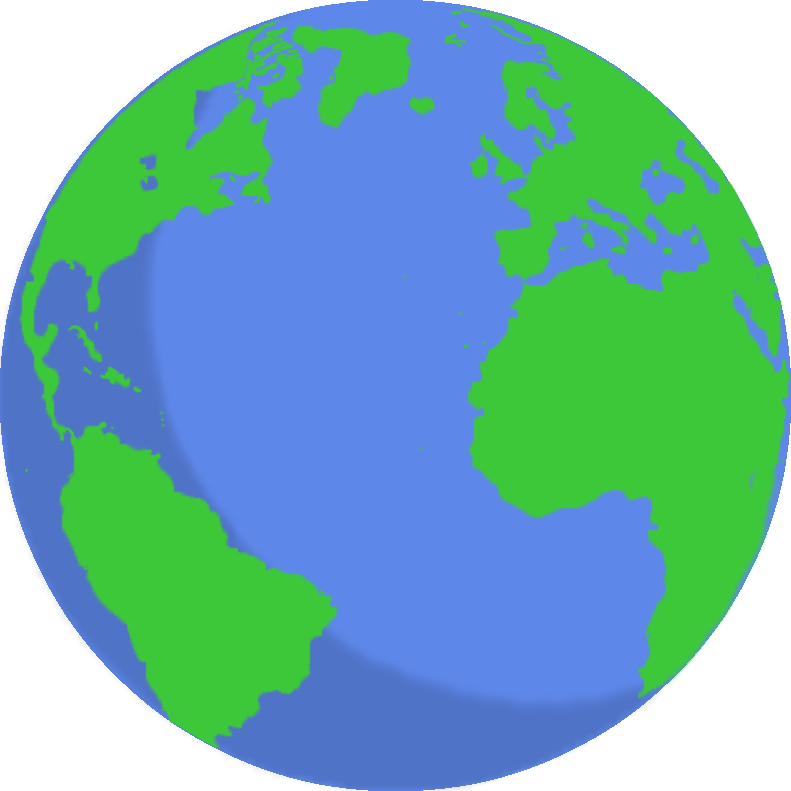}};
\draw[yellow,thick, densely dashed] (2,-.5) arc (-90:90:1cm and 2.5cm);
\draw[yellow,thick, densely dashed] (2,-.5) arc (-90:90:2cm and 2.5cm);
\draw[yellow,thick, densely dashed] (2,-.5) arc (-90:90:-1.5cm and 2.5cm);
\draw[yellow,thick, densely dashed] (2,-.5) arc (-90:90:0.5cm and 2.5cm);
\draw[yellow,thick, densely dashed] (2,-.5) arc (-90:90:1.5cm and 2.5cm);
\draw[yellow,thick, densely dashed] (2,-.5) arc (-90:90:-2cm and 2.5cm);
\draw[yellow,thick, densely dashed] (2,-.5) arc (-90:90:-1cm and 2.5cm);
\draw[yellow,thick, densely dashed] (2,-.5) arc (-90:90:-0.5cm and 2.5cm);
\draw[yellow,thick, densely dashed] (2,-.5) arc (-90:90:0cm and 2.5cm);

\draw [-stealth][black,very thick](2-1.33,3.15) -- (2-.94,3.07);
\draw [-stealth][black,very thick](2-.94,3.07) -- (2-.48,3);
\draw [-stealth][black,very thick](2-.48,3) -- (2,3);
\draw [-stealth][black,very thick](2,3) -- (2.48,3);
\draw [-stealth][black,very thick](2.48,3) -- (2.94,3.07);
\draw [-stealth][black,very thick](2.94,3.07) -- (2.99,2) circle (2pt);

\draw [-stealth][white,very thick](2-1.33,3.15) -- (2-1.5,2);
\draw [-stealth][white,very thick](2-1.5,2) -- (2-.99,2);
\draw [-stealth][white,very thick](2-.99,2) -- (2,2);
\draw [-stealth][white,very thick](2,2) -- (2.48,2);
\draw [-stealth][white,very thick](2.48,2) -- (2.99,2);

\filldraw[red] (2-1.53,3) circle (1pt) node[anchor= east]{$1$};
\draw [red,very thick](2-1.33,3.15) -- (2-1.53,3);
\filldraw[red] (2.79,2.3) circle (1pt) node[anchor= south]{$2$};
\draw [red,very thick](2.99,2) -- (2.79,2.3);

\filldraw[yellow] (2,1) circle (2pt);
\filldraw[yellow] (2,2) circle (2pt);
\filldraw[yellow] (2,3) circle (2pt);
\filldraw[yellow] (2,4) circle (2pt);
\filldraw[yellow] (2,0) circle (2pt);

\filldraw[yellow] (2.45,1) circle (2pt);
\filldraw[yellow] (2.48,2) circle (2pt);
\filldraw[yellow] (2.48,3) circle (2pt);
\filldraw[yellow] (2.3,4) circle (2pt);
\filldraw[yellow] (2.3,0) circle (2pt);

\filldraw[yellow] (2-.45,1) circle (2pt);
\filldraw[yellow] (2-.48,2) circle (2pt);
\filldraw[yellow] (2-.48,3) circle (2pt);
\filldraw[yellow] (2-.3,4) circle (2pt);
\filldraw[yellow] (2-.3,0) circle (2pt);

\filldraw[yellow] (2.6,-.05) circle (2pt);
\filldraw[yellow] (2.93,1-.07) circle (2pt);
\filldraw[yellow] (2.99,2) circle (2pt) node[anchor= north]{$S_2$};
\filldraw[yellow] (2.94,3.07) circle (2pt);
\filldraw[yellow] (2.6,4.05) circle (2pt);

\filldraw[yellow] (2-.6,-.05) circle (2pt);
\filldraw[yellow] (2-.93,1-.07) circle (2pt);
\filldraw[yellow] (2-.99,2) circle (2pt);
\filldraw[yellow] (2-.94,3.07) circle (2pt);
\filldraw[yellow] (2-.6,4.05) circle (2pt);

\filldraw[yellow] (2.85,-.1) circle (2pt);
\filldraw[yellow] (3.33,1-.15) circle (2pt);
\filldraw[yellow] (3.5,2) circle (2pt);
\filldraw[yellow] (3.33,3.15) circle (2pt);
\filldraw[yellow] (2.9,4.1) circle (2pt);

\filldraw[yellow] (2-.85,-.1) circle (2pt);
\filldraw[yellow] (2-1.33,1-.15) circle (2pt) ;
\filldraw[yellow] (2-1.5,2) circle (2pt);
\filldraw[yellow] (2-1.33,3.15) circle (2pt) node[anchor= south]{$S_1$};
\filldraw[yellow] (2-.9,4.1) circle (2pt);

\filldraw[yellow] (3.03,-.18) circle (2pt);
\filldraw[yellow] (3.72,1-.3) circle (2pt);
\filldraw[yellow] (4,2) circle (2pt);
\filldraw[yellow] (3.76,3.25) circle (2pt);
\filldraw[yellow] (3.04,4.14) circle (2pt);

\filldraw[yellow] (2-1.03,-.18) circle (2pt);
\filldraw[yellow] (2-1.72,1-.3) circle (2pt);
\filldraw[yellow] (0,2) circle (2pt);
\filldraw[yellow] (2-1.76,3.25) circle (2pt);
\filldraw[yellow] (2-1.04,4.14) circle (2pt);

\end{tikzpicture}
\caption{\label{fig:constellation} A satellite quantum internet constellation with $N$ polar orbits with $M$ satellites in each orbit. The constellation distributes entanglement between points $1$ and $2$ via black or white satellite paths. }
\end{figure}

Next we consider a full Walker constellation \cite{walker} with $N$ equispaced polar orbits with $M$ satellites in each orbit, as illustrated in Fig. \ref{fig:constellation}. A subset of the constellation comprises EPSs, while the remaining are 4-port RSs with ISLLs to each nearest-neighbor satellite. This is not meant to be an optimal choice for an internet constellation, but we choose this configuration for its simplicity as an illustrative case. Inspired by studies on Starlink optical routing \cite{Handley}, we consider a simple routing protocol for the entanglement distribution between two OGSs. At any time, an EPS can transmit photons to four nearest-neighbor satellites on the constellation. As shown in Fig. \ref{fig:constellation}, OGS1 and OGS2 find the closest satellites $S_1$ and $S_2$ in the constellation. Consider black and white paths between $S_1$ and $S_2$. For each path, the middle satellite emits photons, and the path with the highest expected distribution rate is selected. We consider satellites equipped with 10 GHz entangled photon sources and telescope apertures with $r_a=0.25$ cm and $0.35$ cm. Our simulations include $20\%$ pointing errors and $1\%$ loss for each relay satellite. Tables II and III report the maximum achievable ebit distribution rates between cities across the US, Europe, and Asia, for $r_a=0.25$ cm and $0.35$ cm, respectively. We vary the constellation size with $N=M$ to maximize the achievable rate. Interestingly, for most locations, entanglement distribution rates of a few MHz are achievable with a midsize constellation.

\begin{table}[h!]
  \begin{center}
    \label{tab:table2}
    \footnotesize
    \begin{tabular}{l|l|l} 
      \textbf{OGS1-OGS2} &  $N\times M$ & rate\\
      \hline
      Los Angeles-New York & 5$\times $5 & 5.5 MHz\\
      London-New York  & 8$\times $8 & 0.2 MHz\\
      Berlin-New York  & 4$\times $4 & 0.1 MHz\\
      London-San Francisco  & 8$\times $8 & 0.6 MHz\\
      Los Angeles-Tokyo & 5$\times $5 & 0.3 MHz\\
      Los Angeles-Delhi & 6$\times $6 & 0.4 MHz \\
      San Francisco-Doha & 6$\times $6 & 0.3 MHz\\
      
    \end{tabular}
  \end{center}
      \caption{Maximum entanglement distribution rate between different cities with a polar-orbit constellation with nearest-neighbor ISLLs. Satellites are equipped with a 10GHz source and telescopes with aperture radius 25 cm.}
\end{table} 

\begin{table}[h!]
  \begin{center}
    \label{tab:table2}
    \footnotesize
    \begin{tabular}{l|l|l} 
      \textbf{OGS1-OGS2} &  $N\times M$ & rate\\
      \hline
      Los Angeles-New York & 5$\times $5 & 26 MHz\\
      London-New York  & 8$\times $8 & 3.6 MHz\\
      Berlin-New York  & 7$\times $7 & 1.4 MHz\\
      London-San Francisco  & 8$\times $8 & 1.0 MHz\\
      Los Angeles-Tokyo & 5$\times $5 & 1.9 MHz\\
      Los Angeles-Delhi & 6$\times $6 & 2.5 MHz \\
      Los Angeles-Doha & 6$\times $6 & 1.1 MHz\\
      
    \end{tabular}
  \end{center}
      \caption{Maximum entanglement distribution rate between different cities with a polar-orbit constellation with nearest-neighbor ISLLs. Satellites are equipped with a 10GHz source and telescopes with aperture radius 35 cm.}
\end{table} 

\section{Discussion}

We demonstrate the possibility of building a global quantum internet with a constellation of LEO satellites with payloads carrying entangled-photon pair sources and mirror-based relay satellites. Our proof-of-concept study can be extended to include more error models that capture additional sources of noise and loss, and the constellation can be optimized to maximize the achievable entanglement distribution rates. Nonetheless, with the current model, we demonstrate that entanglement distribution rates of few MHz across the globe are achievable using existing quantum source and satellite technologies.

Going forward, several additional challenges must also be considered. For example, background sunlight presents a major challenge for daytime operation of a satellite quantum network. Operating at less sensitive wavelengths may be one way to circumvent this limitation. Additionally, the use of temporal and spatial filtering techniques can help reduce the background light arriving at the detectors, that has been modeled and shown to increase the signal-to-noise ratio during daytime operation \cite{gruneisen2021adaptive}. Also, note that entanglement is typically an offline resource in quantum protocols; thus, in principle, one can distribute quantum signals to OGSs at night and store them in quantum memories for daytime use.

Other major challenges in satellite-based quantum networking are air turbulence and cloud coverage. Adaptive optics techniques with fast tip-tilt and mirror deformation capabilities can compensate wavefront errors, which can significantly improve quantum communication rates \cite{gruneisen2021adaptive}. For a single satellite, cloud coverage would prevent the establishment of a communications link between the satellite and OGSs. One practical solution is to install redundant OGSs in multiple locations to bypass the blockage; however, more exotic solutions have also been proposed to fundamentally solve the cloud problem, including cloud-piercing techniques \cite{schimmel2018free} or quantum control of cloud molecules to produce acoustic waves to open a clear channel in the clouds \cite{schroeder2020molecular}.

In this work, we chose the operation wavelength of 810 nm to maximize the performance of the designed quantum internet. We repeated our simulations for the telecom wavelength of 1550 nm and found a 50\%-90\% reduction in the entanglement distribution rates. This may imply that 810 nm should be the preferred design choice; however, this wavelength comes with its own challenges. To connect an 810 nm satellite network to a regional fiber-based quantum network that operates at telecom wavelengths, we need frequency conversion, which adds to the total network loss. Another aspect is leveraging the existing satellite lasercom technology, which currently operates at 1550 nm. These systems, including the laser terminal and ground telescope, cannot be readily used at 810 nm and require design changes, including coatings and geometric parameters. Therefore, if cost is a factor, using 1550 nm could be the right choice. Another advantage of 1550 nm is that current photonic integrated circuit platforms generate higher rates of entangled photon pairs at this wavelength \cite{Moody2025}. Thus, the decision about the wavelength needs to be made through a rigorous design analysis considering all technical and economic factors.

In conclusion, we discuss how to improve our proposed design to achieve gigahertz rates of entanglement distribution. One clear path involves further enhancing the rate of the entanglement generation source, which could be realized using multiplexing photonic integrated circuits. Another important direction is optimizing our constellation size and architecture, which is by no means currently optimal. We expect that optimizing the constellation architecture can also enhance the rate by at least an order of magnitude. For example, the global population and urban areas are not distributed uniformly, with a majority residing in 30-50 degree latitudes. Current internet constellations, such as Starlink, employ a mixed architecture with satellites at different altitudes and inclinations \cite{cakaj}. Another extension of our proposal is to go beyond nearest-neighbor satellite links and allow for shorter paths for the photons to travel \cite{Chaudhry21}. Optimal constellation design and routing protocols are some future directions for research on satellite-based quantum networking. Lastly, enhanced pointing accuracy for laser links can significantly boost the performance of quantum satellite constellations. In this work, we considered a 20$\%$ error rate. Advancing the technology to achieve highly precise pointing with below 5 $\%$ error is another requirement for realizing a satellite-based global quantum internet \cite{song2017}.\\


\section{Acknowledgment}

\noindent We thank Mostafa Abasifard for discussions on the experimental challenges of daytime satellite quantum communication.

\bibliography{shabanibib}

\end{document}